# Exact solutions for time-optimal control of spin I=1 by NMR

**Vitaly Shauro**

Kirensky Institute of Physics, Russian Academy of Sciences, Siberian Branch,
Akademgorodok 50, bld. 38, Krasnoyarsk, Russia

E-mail: rsa@iph.krasn.ru

**Abstract** We consider the problem of time-optimal control of quadrupole nucleus with the spin $I$=1 by NMR. In contrast to the conventional methods based on selective pulses, the control is implemented using nonselective pulses separated by free evolution intervals. Using the Cartan decomposition, the system of equations is obtained for finding parameters of a control field. Partial time-optimal solutions for the important single-qutrit gates (selective rotations and quantum Fourier transform) are found. The strong dependence of minimum gate implementation times on global phase of the gate is observed. The analytical values of minimum times are consistent with the numerical data.

1. Introduction

Over the past two decades, the existing methods for quantum system control have been significantly improved in respect of both their theory and experimental implementations [1−3]. The research in this direction is of high practical importance, since the problems solved ranges from the control of chemical reactions to the development of quantum computers and quantum communication channels.

One of the most interesting and promising area that requires the high-precision quantum state control is the quantum information processing [4, 5]. As is known, for the successful implementation of quantum algorithms the error of elementary logic operations (gates) must not exceed a certain threshold value. To meet this requirement during experiments, the quantum system dynamics should be effectively controlled. In addition to high fidelity of manipulations made, they must take minimum time to reduce relaxation effects. Moreover, estimation of the minimum (critical) time of transformations in specific systems is directly related to the efficiency of execution of quantum algorithm [6, 7].

Unfortunately, the available analytical approaches to the estimation of the minimum time and determination of optimal control conditions are extremely hard to apply to quantum systems with a large number of states. Currently, only few simple models comprising several qubits represented by spins 1/2 allow finding exact analytical solutions of the optimal control problem [8−11]. Therefore, numerical methods are often used [12-14] for finding the optimal control by a system to establish a desired quantum state or, in the general case, a certain unitary evolution of a quantum system.

Already in the case of three-level quantum elements, qutrits [15−17], it is difficult to find the optimal ways of universal gate implementation. Meanwhile, the qutrits are the most interesting multilevel systems. They ensure the fastest growth of the Hilbert space dimension with increasing number of logical elements [18] and are more effective than qubits in quantum communications [19]. Therefore, many researchers explore implementation of ternary quantum computations, both in quantum circuits for solving various problems [16, 17, 20, 21] and in qutrit gates based on different physical models [20−26].

In this work, we solve the problem of the time-optimal control for a single quadrupole nucleus with the spin $I=1$ by the nuclear magnetic resonance (NMR) technique. The quantum states with different projections of the spin $I=1$ are considered as the basic qutrit states. Instead of the traditional methods that use weak selective radio-frequency (rf) pulses [23, 26, 27] for exciting individual energy transitions in a multilevel system, we use the strong rf control field that affects all states of a system simultaneously (the so-called nonselective excitation) [27]. Using the Cartan decomposition [8], we managed to obtain the general equation for finding parameters of a control field to implement an arbitrary qutrit gate. We obtained partial time-optimal solutions for selective rotation gates and a quantum Fourier transform (QFT) with regard to the global phase of gates.

In Sec. II, we describe the model of a quadrupole nucleus with the spin $I=1$ controlled by NMR. Using the Cartan decomposition, the general equation is derived for solving the time-optimal control problem for this model. In Sec. III, partial solutions of this equation for the selective rotation gates and QFT are presented and the effect of the global phase on the minimum gate implementation time is discussed. Section IV contains the conclusions.

## 2. Optimal control for spin I=1 by NMR

Let us consider a quadrupole nucleus with the spin $I=1$ in a strong static magnetic field and a control rf magnetic field. In the reference frame rotating around the static field direction (axis $z$) at rf field frequency $\omega_{rf}$, the Hamiltonian of our model acquires the form [27]

$$H(t) = (\omega_{rf} - \omega_0)I_z + H_q + u_x(t)I_x + u_y(t)I_y, \qquad H_q = q\left(I_z^2 - \tfrac{2}{3}\right). \tag{1}$$

Here, $\omega_0$ is the Larmor frequency, $I_\alpha$ is the operator of spin projection onto the axis $\alpha$ ($\alpha = x, y, z$), $H_q$ is the quadrupole interaction of the nucleus with the axially symmetric crystal field gradient, $q$ is the interaction constant, and $u_\alpha(t)$ is the projection of the control rf field onto the axis $\alpha$. Hereinafter, the energy is measured in frequency units with $\hbar = 1$. In addition, we pass to dimensionless time and frequency expressed in units $1/q$ and $q$, respectively. In the absence of the rf field, system (1) has three nonequidistant energy levels for the states with the different values of spin projection $I_z = 0, \pm 1$. We choose these states as a qutrit computational basis.

The system with Hamiltonian (1) is fully controllable in group $SU(3)$ [28]; i.e., for sufficiently large time $T$, there exists such control $u_\alpha(t)$ that allows us to obtain any operator $U_G = U(T) \in SU(3)$ where

$$U(T) = \hat{T} \exp\left(-i\int_0^T H(t)dt\right) \tag{2}$$

is the system evolution operator and $\hat{T}$ is the time-ordering operator.

To find the desired control and minimum time required for implementing some important qutrit gates, we use the Cartan decomposition [8]. First, we need to choose eight generators of algebra $\mathfrak{su}(3)$. The generators can be chosen in different ways and we should select the most convenient set for our task. For example, in [21, 29], the Gell−Mann matrices were considered as $\mathfrak{su}(3)$ generators. In terms of NMR, these generators can be compared with the selective ("soft") pulses affecting individual transitions in a three-level system. To implement this control method, the rf field frequency should be equal to the resonant frequency of one of the allowed transitions ($\omega_{rf} = \omega_0 \pm q$) and the pulse amplitude should meet the condition $|u| \ll q$ [23, 24, 27]. Using a certain sequence of pulses applied at the different transitions, one may implement any unitary system evolution [16, 17]; however, such a control is not time-optimal, since the control field is strictly limited. In view of this, we will consider the control with the use of nonselective ("hard") rf pulses, when $\omega_{rf} = \omega_0$ and $|u| \gg q$. For example, in simplest case of square-shaped pulse along $x$ axis ($x$-pulse) we set the pulse amplitudes $u_x = \Omega \gg q$ and $u_y = 0$ in (1) and corresponding propagator can be expressed as

$$U(t) = \exp[-it(H_q + \Omega I_x)] \approx \exp(-i\theta I_x) = \tfrac{1}{2}\begin{bmatrix} 1+\cos\theta & -i\sqrt{2}\sin\theta & \cos\theta - 1 \\ -i\sqrt{2}\sin\theta & 2\cos\theta & -i\sqrt{2}\sin\theta \\ \cos\theta - 1 & -i\sqrt{2}\sin\theta & 1+\cos\theta \end{bmatrix}, \tag{3}$$

where $\theta = \Omega t$ is angle of nonselective rotation of the spin and $t$ is the pulse length. Similarly, for $y$-pulse the propagator is

$$U(t) = \exp[-it(H_q + \Omega I_y)] \approx \exp(-i\theta I_y) = \tfrac{1}{2}\begin{bmatrix} 1+\cos\theta & -\sqrt{2}\sin\theta & 1-\cos\theta \\ \sqrt{2}\sin\theta & 2\cos\theta & -\sqrt{2}\sin\theta \\ 1-\cos\theta & \sqrt{2}\sin\theta & 1+\cos\theta \end{bmatrix}. \quad (4)$$

As will be seen below, the sequence of nonselective rotations (3)-(4) and free evolution intervals with propagator

$$U(t) = \exp(-itH_q) \quad (5)$$

allows to obtain any system evolution (2). For finding such sequence, it is convenient to take operators $I_x, I_y, I_z, H_q$ as generators and the adjoint operators [8]

$$Ad_k(H_q) = e^{-ik} H_q e^{ik}, \quad k \in \{I_x, I_y\}. \quad (6)$$

As follows from the properties of exponential operators, the evolution under the action of generators (6) is realized by means of two nonselective pulses (3) or (4) separated by an interval of free evolution (5). This can be used to build the pulse sequences for specific gates $U_G$. Thus, as the basis of algebra $\mathfrak{su}(3)$, we choose the following set of matrices

$$L_1 = I_x = \frac{1}{\sqrt{2}}\begin{bmatrix} 0 & 1 & 0 \\ 1 & 0 & 1 \\ 0 & 1 & 0 \end{bmatrix}, \quad L_2 = I_y = \frac{1}{\sqrt{2}}\begin{bmatrix} 0 & -i & 0 \\ i & 0 & -i \\ 0 & i & 0 \end{bmatrix},$$

$$L_3 = I_z = \begin{bmatrix} 1 & 0 & 0 \\ 0 & 0 & 0 \\ 0 & 0 & -1 \end{bmatrix}, \quad L_4 = H_q = \frac{q}{3}\begin{bmatrix} 1 & 0 & 0 \\ 0 & -2 & 0 \\ 0 & 0 & 1 \end{bmatrix},$$

$$L_5 = e^{-i\frac{\pi}{4}I_x} H_q e^{i\frac{\pi}{4}I_x} = \frac{q}{12\sqrt{2}}\begin{bmatrix} \sqrt{2} & 6i & -3\sqrt{2} \\ -6i & -2\sqrt{2} & -6i \\ -3\sqrt{2} & 6i & \sqrt{2} \end{bmatrix},$$

$$L_6 = e^{-i\frac{\pi}{4}I_y} H_q e^{i\frac{\pi}{4}I_y} = \frac{q}{12\sqrt{2}}\begin{bmatrix} \sqrt{2} & 6 & 3\sqrt{2} \\ 6 & -2\sqrt{2} & -6 \\ 3\sqrt{2} & -6 & \sqrt{2} \end{bmatrix},$$

$$L_7 = e^{-i\frac{\pi}{2}I_y} H_q e^{i\frac{\pi}{2}I_y} = \frac{q}{6}\begin{bmatrix} -1 & 0 & 3 \\ 0 & 2 & 0 \\ 3 & 0 & -1 \end{bmatrix}, \quad L_8 = e^{i\frac{\pi}{2}I_y} L_5 e^{-i\frac{\pi}{2}I_y} = \frac{q}{6}\begin{bmatrix} -1 & 0 & -3i \\ 0 & 2 & 0 \\ 3i & 0 & -1 \end{bmatrix}. \quad (7)$$

Next, to find the time-optimal pulse sequence, we use the mathematical tools outlined in [8]. The Cartan decomposition of algebra $\mathfrak{su}(3)$ is

$$\mathfrak{su}(3) = \mathfrak{p} \oplus \mathfrak{k}, \quad (8)$$

where subalgebras $\mathfrak{p}$ and $\mathfrak{k}$ satisfy the commutation relations

$$[\mathfrak{k},\mathfrak{k}] \subset \mathfrak{k}, \quad [\mathfrak{p},\mathfrak{k}] \subset \mathfrak{p}, \quad [\mathfrak{p},\mathfrak{p}] \subset \mathfrak{k}. \quad (9)$$

For the algebra with basis (7), we have

$$\mathfrak{k} = \text{span}\{L_m \mid m = 1, 2, 3\},$$
$$\mathfrak{p} = \text{span}\{L_m \mid m = 4,..,8\}, \qquad (10)$$

where $\text{span}\{L_m\}$ denotes the space spanned by operators $L_m$. If subspace $\mathfrak{h} \subset \mathfrak{p}$ is maximal Abelian subalgebra (Cartan subalgebra), then arbitrary evolution operator (2) can be written in the form [8]

$$U(T) = Q_1 \exp(-i\mathfrak{h}) Q_2, \quad Q = \{e^{-ik} \mid k \in \mathfrak{k}\}. \qquad (11)$$

We take the subspace $\mathfrak{h} = \text{span}\{L_4, L_7\}$ spanned by a pair of commuting generators as Cartan subalgebra. (The Cartan subalgebra $\mathfrak{h}' = \text{span}\{L_4, L_8\}$ can also be taken, but in most cases this choice will complicate the solutions obtained in Sec. III).

Since subalgebra $\mathfrak{k}$ consists of only the spin projection operators, an arbitrary element of the subgroup generated by this subalgebra can be represented in the form of the Euler angles decomposition

$$Q = e^{-i\alpha I_x} e^{-i\beta I_y} e^{-i\gamma I_x}. \qquad (12)$$

In general case, in Eq. (12) one can select two any orthogonal rotation axes. However, for simplicity, we use a series of rotations $x$-$y$-$x$, as in Eq. (12), or $y$-$x$-$y$ at an appropriate permutation of the rotation axes. Thus, for the series $x$-$y$-$x$, Eq. (11) acquires the form

$$U(T) = e^{-i\alpha_1 I_x} e^{-i\beta_1 I_y} e^{-i\gamma_1 I_x} e^{-it_1 L_4} e^{-it_2 L_7} e^{-i\alpha_2 I_x} e^{-i\beta_2 I_y} e^{-i\gamma_2 I_x}. \qquad (13)$$

Equating the right-hand side of Eq. (13) to gate $U_G$ to be implemented, we obtain a system of equations for finding parameters $\alpha_{1,2}, \beta_{1,2}, \gamma_{1,2}, t_{1,2}$. According to the theorems in [8], the solutions of this system of equations are time-optimal when $t_{1,2} > 0$ and the sum $T = t_1 + t_2$ takes the smallest value, $T_{\min}$. Here, rotations (12) are assumed to be implemented for a negligible time using the control field with large amplitude $|u| \gg q$ [pulse length $t \sim 1/\Omega$ in (3)-(4) is much less than duration of free evolution $t \sim 1/q$ in (5)].

After explicit substitution of operators $L_4$ and $L_7$, the product of operators in Eq. (13) represents a ready-to-use method for implementing single-qutrit gates by using a sequence of strong nonselective pulses (actually, from 2 to 8 pulses, depending on the implemented gates as will be shown in Sec III) separated by two intervals of free evolution with durations $t_1$ and $t_2$. The schematic representation of the sequence shown in Figure 1. In contrast to the previously proposed sequences [30, 31], no additional cycling of the pulse sequence to reduce the gate error is required. The errors caused by a finite length of the nonselective pulses [i.e. approximations (3)-(4)] in such sequences were estimated in [31] and the method for their reduction was proposed.

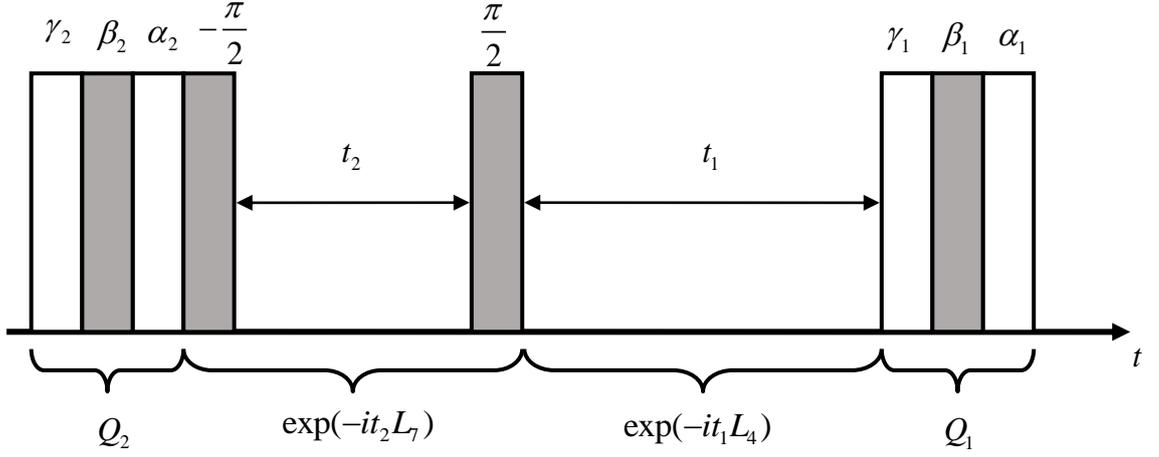

**Fig. 1.** The schematic representation of the pulse sequence (13). White and gray squares denote nonselective *x*- and *y*-pulses respectively. The angle of spin rotation is shown above corresponding pulses. Note that the order of pulses in time is opposite to the order of factors in (13) according to the traditional NMR notations.

### 3. Exact solutions for selective rotation gates and QFT

In the general case, the search for parameters $\alpha_{1,2}, \beta_{1,2}, \gamma_{1,2}, t_{1,2}$ in Eq. (13) is a complex problem. However, it can always be solved numerically. If gate $U_G$ is sufficiently simple, the analysis of numerical solutions often allows one to guess the values of some parameters, because they take simple values. This makes the problem much simpler, since we can now find analytical expressions for the rest parameters and, thus, obtain some exact time-optimal solutions. Note that due to the periodicity of the factors in Eq. (13) and the fact that the same rotation $Q$ can be obtained using different rotation axes in Eq. (12), there exist many solutions of the control problem, including suboptimal ($T > T_{\min}$) and unphysical ($t_{1,2} < 0$) ones. Therefore, we present only a few simple partial time-optimal solutions for the most important gates used in quantum algorithms based on qutrits [15−17, 21]. We consider the operators of selective rotation around the axes *x* and *y* by angle $\theta$,

$$R_x^{1-2}(\theta) = \begin{bmatrix} \cos\frac{\theta}{2} & i\sin\frac{\theta}{2} & 0 \\ i\sin\frac{\theta}{2} & \cos\frac{\theta}{2} & 0 \\ 0 & 0 & 1 \end{bmatrix}, \quad R_y^{1-2}(\theta) = \begin{bmatrix} \cos\frac{\theta}{2} & -\sin\frac{\theta}{2} & 0 \\ \sin\frac{\theta}{2} & \cos\frac{\theta}{2} & 0 \\ 0 & 0 & 1 \end{bmatrix},$$

$$R_x^{2-3}(\theta) = \begin{bmatrix} 1 & 0 & 0 \\ 0 & \cos\frac{\theta}{2} & i\sin\frac{\theta}{2} \\ 0 & i\sin\frac{\theta}{2} & \cos\frac{\theta}{2} \end{bmatrix}, \quad R_y^{2-3}(\theta) = \begin{bmatrix} 1 & 0 & 0 \\ 0 & \cos\frac{\theta}{2} & -\sin\frac{\theta}{2} \\ 0 & \sin\frac{\theta}{2} & \cos\frac{\theta}{2} \end{bmatrix}, \quad (14)$$

$$R_x^{1-3}(\theta) = \begin{bmatrix} \cos\frac{\theta}{2} & 0 & i\sin\frac{\theta}{2} \\ 0 & 1 & 0 \\ i\sin\frac{\theta}{2} & 0 & \cos\frac{\theta}{2} \end{bmatrix}, \quad R_y^{1-3}(\theta) = \begin{bmatrix} \cos\frac{\theta}{2} & 0 & -\sin\frac{\theta}{2} \\ 0 & 1 & 0 \\ \sin\frac{\theta}{2} & 0 & \cos\frac{\theta}{2} \end{bmatrix},$$

and the quantum Fourier transform gate

$$F = \frac{1}{\sqrt{3}}\begin{pmatrix} 1 & 1 & 1 \\ 1 & \sigma & \sigma^2 \\ 1 & \sigma^2 & \sigma \end{pmatrix}, \qquad \sigma = \exp\left(\frac{2\pi i}{3}\right). \tag{15}$$

Since these operators are defined in group $U(3)$, for the model with traceless Hamiltonian (1) the gates can be implemented accurate to the common phase factor [2, 6]; i.e., at $U_G \in U(3)$, we should solve the system of equations [in the case of the rotations $x$-$y$-$x$ in (12)]

$$e^{i\phi}U_G = e^{-i\alpha_1 I_x} e^{-i\beta_1 I_y} e^{-i\gamma_1 I_x} e^{-it_1 L_4} e^{-it_2 L_7} e^{-i\alpha_2 I_x} e^{-i\beta_2 I_y} e^{-i\gamma_2 I_x}. \tag{16}$$

In general case, the global phase $\phi$ can be chosen from the set of values

$$\phi = \phi_0 + 2\pi p/N, \quad p = 0, 1, ..., N-1, \tag{17}$$

where $N$ is the Hilbert space dimension and $\phi_0$ is the smallest angle $\phi_0 \in [0, \pi]$ at which $\det(e^{i\phi_0}U_G) = 1$ [6]. For qutrit ($N=3$) the global phase can take the values $\phi = 0, \frac{2\pi}{3}, \frac{4\pi}{3}$ for selective rotations (14) and $\phi = \frac{\pi}{6}, \frac{5\pi}{6}, \frac{9\pi}{6}$ for QFT (15). As is known, the minimum gate implementation time depends on the global phase value [6, 14, 32, 33]. Such dependence for model (1) was numerically studied in [32] for the QFT gate and can be formally explained by multivaluedness of the logarithm function of complex numbers.

Partial time-optimal solutions of Eq. (16) with different global phases for selective rotation gates (14) and QFT (15) are given in Tables 1 and 2, respectively. Figure 2 shows the dependences of the minimum time of selective rotation gate implementation on the angle of rotation for the solutions with different global phases. It can be seen that the dependences are qualitatively different. For the global phase $\phi = 0$, the normal behavior is observed, where the minimum time increases with the rotation angle. For $\phi = \frac{4\pi}{3}$, the situation is qualitatively different. Moreover, for $\phi = \frac{2\pi}{3}$, the minimum time $T_{\min} = \pi$ is independent of the angle of rotation. In the case of rotation $R^{1-3}(\theta)$, which corresponds to the forbidden transition according to the selection rules, the lines intersect at $\theta = \frac{2\pi}{3}$.

Note that if we decompose the value of $T_{min}$ for $R^{1-2}(\theta)$ with $\phi = 0$ (Table 1) at small angles of rotation, we obtain

$$T_{min} = 3\arccos(\cos^2 \tfrac{\theta}{4}) = 6\arcsin(\tfrac{1}{\sqrt{2}}\sin\tfrac{\theta}{4}) \approx \tfrac{6}{\sqrt{2}}\sin\tfrac{\theta}{4} \approx \tfrac{3\theta}{2\sqrt{2}}. \qquad (18)$$

This value obtained in [30] using the approximate approach is slightly different from the numerical data for the angles $\theta \sim \pi$ [34]. The exact analytical values of $T_{min}$ given in Table 1 are consistent with the numerical estimates (Fig. 2) obtained by us with the use of the optimization procedure described in [32]. For the QFT gate, the consistence of the data with the numerical estimates from [32] is also observed.

## 4. Conclusions

Using the Cartan decomposition, we obtained the exact analytical solution of the time-optimal control problem for a quadrupole nucleus with the spin $I$=1 controlled by a strong nonselective rf field. The minimum times required for implementing the important single-qutrit gates (selective rotations and QFT) were found. The extraordinary dependence of the minimum time for the selective rotation gate implementation on the angle of rotation was observed at some global phase values. This demonstrates the importance of taking the global phase into account in solving the optimal control problem for quantum information processing [2, 6, 14, 32, 33]. All analytical results are consistent with the data obtained using numerical methods of the control theory [32, 34].

The presented analytical expressions give a practical method for controlling the spin $I$=1 using a sequences of short nonselective pulses with two intervals of free evolution. This control technique can be used, for example, in NMR experiments with liquid crystals [23, 26], where the quadrupole interaction is weakened by the fast rotation of molecules, which restricts application of weak selective pulses.

Unfortunately, the approach used cannot be directly applied to the large spins $I$ >1, because in this case the commutation relations (9) are violated. It is still possible to obtain some pulse sequences for the implementation simple qudit gates [31], but we cannot assert that the resulting solutions are time-optimal.

**Table 1.** Partial time-optimal solutions of Eq. (16) for implementing selective rotation gates (14) with different global phases. The asterisk denotes the solution in which operator $L_7$ in (16) was replaced by $L_8$ for simplicity. We denote $\xi = \frac{1}{2}\arctan\left(\dfrac{2\sqrt{2}\sin\frac{\theta}{2}}{1+3\cos\frac{\theta}{2}}\right)$ and $\eta = \frac{\pi}{2}+\xi$.

| | $\alpha_1$ | $\beta_1$ | $\gamma_1$ | $\alpha_2$ | $\beta_2$ | $\gamma_2$ | $t_1$ | $t_2$ | Q | $T_{\min}$ and parameters |
|---|---|---|---|---|---|---|---|---|---|---|
| **1) Global phase $\phi=0$** | | | | | | | | | | |
| $R_x^{1-2}$ | $\xi$ | $-\frac{\pi}{4}$ | $\frac{\pi}{2}$ | $-\frac{\pi}{2}$ | $\frac{\pi}{4}$ | $\xi$ | $\tau$ | $2\tau$ | x-y-x | |
| $R_y^{1-2}$ | $0$ | $\xi$ | $-\frac{\pi}{4}$ | $\frac{\pi}{4}$ | $\xi$ | $0$ | $2\tau$ | $\tau$ | x-y-x | $\tau = \arccos(\cos^2\frac{\theta}{4})$, |
| $R_x^{2-3}$ | $\xi$ | $\frac{\pi}{4}$ | $\frac{\pi}{2}$ | $-\frac{\pi}{2}$ | $-\frac{\pi}{4}$ | $\xi$ | $\tau$ | $2\tau$ | x-y-x | $T_{\min} = 3\arccos(\cos^2\frac{\theta}{4})$ |
| $R_y^{2-3}$ | $0$ | $\xi$ | $\frac{\pi}{4}$ | $-\frac{\pi}{4}$ | $\xi$ | $0$ | $2\tau$ | $\tau$ | x-y-x | |
| $R_x^{1-3}$ | $0$ | $0$ | $0$ | $0$ | $0$ | $0$ | $\tau$ | $2\tau$ | | $\tau = \frac{\theta}{2}$, |
| $^*R_y^{1-3}$ | $0$ | $0$ | $0$ | $0$ | $0$ | $0$ | $\tau$ | $2\tau$ | | $T_{\min} = \frac{3\theta}{2}$ |
| **2) Global phase $\phi=2\pi/3$** | | | | | | | | | | |
| $R_x^{1-2}$ | $\xi$ | $\frac{\pi}{4}$ | $\pi$ | $0$ | $-\frac{\pi}{4}$ | $\xi$ | $\tau_2$ | $\tau_1$ | x-y-x | |
| $R_y^{1-2}$ | $\eta$ | $-\frac{\pi}{2}$ | $\frac{\pi}{4}$ | $\frac{\pi}{4}$ | $\frac{\pi}{2}$ | $\eta$ | $\tau_1$ | $\tau_2$ | y-x-y | $\tau_1 = \pi - \arccos(\cos^2\frac{\theta}{4})$, |
| $R_x^{2-3}$ | $\xi$ | $-\frac{\pi}{4}$ | $\pi$ | $0$ | $\frac{\pi}{4}$ | $\xi$ | $\tau_2$ | $\tau_1$ | x-y-x | $\tau_2 = \arccos(\cos^2\frac{\theta}{4})$, |
| $R_y^{2-3}$ | $\eta$ | $\frac{\pi}{2}$ | $\frac{\pi}{4}$ | $\frac{\pi}{4}$ | $-\frac{\pi}{2}$ | $\eta$ | $\tau_1$ | $\tau_2$ | y-x-y | $T_{\min} = \pi$ |
| $R_x^{1-3}$ | $\frac{\pi}{2}$ | $-\frac{\pi}{2}$ | $0$ | $0$ | $-\frac{\pi}{2}$ | $\frac{\pi}{2}$ | $\tau_1$ | $\tau_2$ | x-y-x | $\tau_1 = \frac{\theta}{2}$, |
| $R_y^{1-3}$ | $-\frac{\pi}{2}$ | $\frac{\pi}{4}$ | $0$ | $0$ | $\frac{\pi}{4}$ | $-\frac{\pi}{2}$ | $\tau_1$ | $\tau_2$ | x-y-x | $\tau_2 = \pi - \frac{\theta}{2}$, $T_{\min} = \pi$ |
| **3) Global phase $\phi=4\pi/3$** | | | | | | | | | | |
| $R_x^{1-2}$ | $\xi$ | $\frac{\pi}{4}$ | $-\frac{\pi}{2}$ | $-\frac{\pi}{2}$ | $\frac{3\pi}{4}$ | $\xi$ | $\tau_1$ | $\tau_2$ | x-y-x | |
| $R_y^{1-2}$ | $\xi$ | $\frac{\pi}{4}$ | $-\frac{\pi}{2}$ | $-\frac{\pi}{2}$ | $-\frac{\pi}{4}$ | $\xi$ | $\tau_1$ | $\tau_2$ | y-x-y | $\tau_1 = \pi - \arccos(\cos^2\frac{\theta}{4})$, |
| $R_x^{2-3}$ | $\xi$ | $-\frac{\pi}{4}$ | $\frac{\pi}{2}$ | $\frac{\pi}{2}$ | $-\frac{3\pi}{4}$ | $\xi$ | $\tau_1$ | $\tau_2$ | x-y-x | $\tau_2 = \pi - 2\arccos(\cos^2\frac{\theta}{4})$, |
| $R_y^{2-3}$ | $\xi$ | $-\frac{\pi}{4}$ | $\frac{\pi}{2}$ | $\frac{\pi}{2}$ | $\frac{\pi}{4}$ | $\xi$ | $\tau_1$ | $\tau_2$ | y-x-y | $T_{\min} = 2\pi - 3\arccos(\cos^2\frac{\theta}{4})$ |
| $R_x^{1-3}$ | $-\frac{\pi}{2}$ | $\frac{\pi}{2}$ | $0$ | $0$ | $\frac{\pi}{2}$ | $-\frac{\pi}{2}$ | $\tau_1$ | $\tau_2$ | y-x-y | $\tau_1 = \pi - \theta$, |
| $R_y^{1-3}$ | $\frac{\pi}{2}$ | $\frac{\pi}{4}$ | $0$ | $0$ | $\frac{\pi}{4}$ | $\frac{\pi}{2}$ | $\tau_1$ | $\tau_2$ | y-x-y | $\tau_2 = \pi - \frac{\theta}{2}$, $T_{\min} = 2\pi - \frac{3\theta}{2}$ |

**Table 2.** Partial time-optimal solutions of Eq. (16) for implementing QFT gate (15) with different global phases.

| $\phi$ | $\alpha_1$ | $\beta_1$ | $\gamma_1$ | $\alpha_2$ | $\beta_2$ | $\gamma_2$ | $t_1$ | $t_2$ | Q | $T_{\min}$ and parameters |
|---|---|---|---|---|---|---|---|---|---|---|
| $\frac{\pi}{6}$ | $-\frac{\pi}{2}$ | $\frac{\pi}{3}$ | $\frac{\pi}{4}$ | $-\frac{\pi}{4}$ | $\frac{2\pi}{3}$ | $-\frac{\pi}{2}$ | $\tau_1$ | $\tau_2$ | x-y-x | $\tau_1 = \pi - 2\arccos(\sqrt{\tfrac{2}{3}})$, $\tau_2 = \pi - \arccos(\sqrt{\tfrac{2}{3}})$, $T_{\min} = 2\pi - 3\arccos(\sqrt{\tfrac{2}{3}})$ |
| $\frac{5\pi}{6}$ | $-\frac{\pi}{2}$ | $\frac{\pi}{3}$ | $-\frac{\pi}{4}$ | $-\frac{\pi}{4}$ | $\frac{\pi}{3}$ | $\frac{\pi}{2}$ | $2\tau$ | $\tau$ | x-y-x | $\tau = \arccos(\sqrt{\tfrac{2}{3}})$, $T_{\min} = 3\arccos(\sqrt{\tfrac{2}{3}})$ |
| $\frac{9\pi}{6}$ | $-\frac{\pi}{2}$ | $\frac{\pi}{6}$ | $\frac{\pi}{4}$ | $-\frac{\pi}{4}$ | $\frac{\pi}{6}$ | $\frac{\pi}{2}$ | $\tau_1$ | $\tau_2$ | x-y-x | $\tau_1 = \arccos(\sqrt{\tfrac{2}{3}})$, $\tau_2 = \pi - \arccos(\sqrt{\tfrac{2}{3}})$, $T_{\min} = \pi$ |

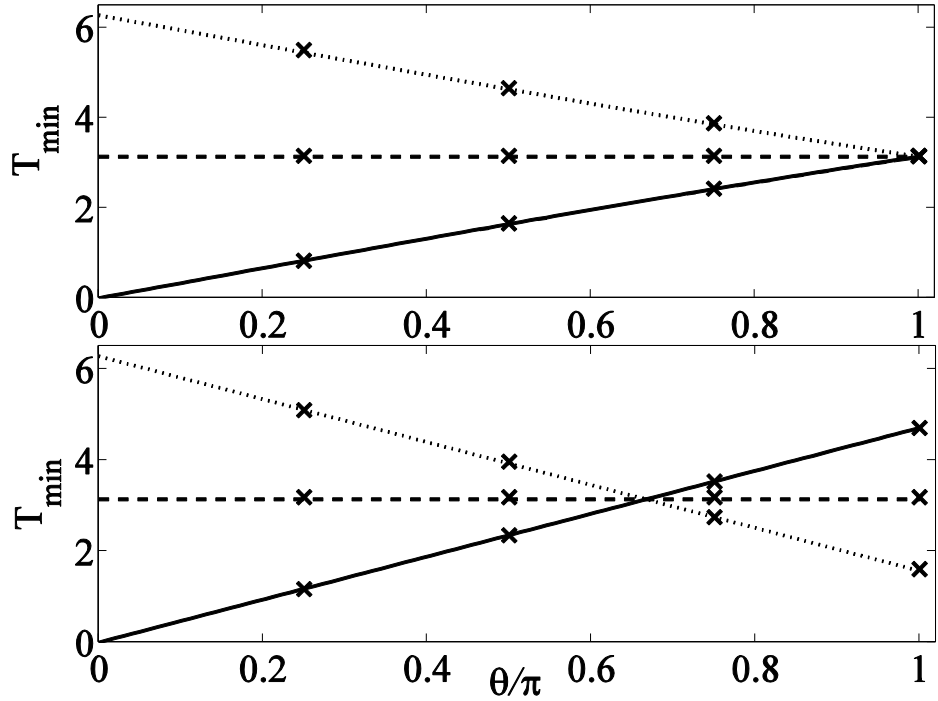

**Fig. 2.** The dependence of minimum implementation time $T_{\min}$ (Table 1) of selective rotation gate (14) on the angle of rotation $\theta$ for the global phases $\phi = 0$ (solid line), $\phi = \frac{2\pi}{3}$ (dashed line), and $\phi = \frac{4\pi}{3}$ (dotted line). The results for rotations $R^{1-2}(\theta)$ or $R^{2-3}(\theta)$ (top) and $R^{1-3}(\theta)$ (bottom) are shown. The crosses shows the estimation of $T_{\min}$ for $\theta = \frac{\pi n}{4}$, $n = 1,...,4$ obtained using the numerical methods described in [32].

**Acknowledgments**

The author thanks V.E. Zobov for fruitful discussions. This study was supported by the Russian Foundation for Basic Research, project no. 14-07-31086.

**References**

[1] Brif, C., Chakrabarti, R., Rabitz, H.: Control of quantum phenomena: past, present and future. New J. Phys. **12,** 075008 (2010)

[2] Jones, J.A.: Quantum computing with NMR. Prog. NMR. Specrosc. **59,** 91(2011)

[3] Wu, R., Zhang, J., Li, Ch., Long, G., Tarn, T.: Control problems in quantum systems. Chines Sci. Bull. **57,** 18 (2012)

[4] Nielsen, M.A., Chuang, I.L.: Quantum Computation and Quantum Information. Cambridge University Press, Cambridge (2000)

[5] Valiev, K.A., Kokin, A.A.: Quantum Computers: Hopes and Reality. Regul. Khaot. Din., Izhevsk (2001)

[6] Schulte-Herbrüggen, T., Spörl, A., Khaneja, N., Glaser, S.J.: Optimal control-based efficient synthesis of building blocks of quantum algorithms: A perspective from network complexity towards time complexity. Phys. Rev. A **72,** 042331 (2005)

[7] Koike, T., Okudaira, Y.: Time complexity and gate complexity. Phys. Rev. A **82,** 042305 (2010)

[8] Khaneja, N., Brockett, R., Glaser, S.J.: Time optimal control in spin systems. Phys. Rev. A **63,** 032308 (2001)

[9] Carlini, A., Hosoya, A., Koike, T., Okudaira, Y.: Time-optimal unitary operations. Phys. Rev. A **75,** 042308 (2007)

[10] Bin, L., ZuHuan, Y., ShaoMing, F., XianQing, L.–J.: Time optimal quantum control of two-qubit systems. Science China G **56,** 2116 (2013)

[11] Yuan, H., Wei, D., Zhang, Y., Glaser, S., Khaneja, N.: Efficient synthesis of quantum gates on indirectly coupled spins. Phys. Rev. A **89,** 042315 (2014)

[12] Khaneja, N., Reiss, T., Kehlet, C., Schulte-Herbrüggen, T., Glaser, S.J.: Optimal control of coupled spin dynamics: design of NMR pulse sequences by gradient ascent algorithms. J. Magn. Reson. **172,** 296 (2005)

[13] Machnes, S., Sander, U., Glaser, S.J., de Fouquieres, P.P., Gruslys, A., Schirmer, S., Schulte-Herbrüggen, T.: Comparing, optimizing, and benchmarking quantum-control algorithms in a unifying programming framework. Phys. Rev. A **84,** 022305 (2011)

[14] Moore Tibbetts, K.W., Brif, C., Grace, M.D., Donovan, A., Hocker, D.L., Ho, T.-S., Wu, R.–B., Rabitz, H.: Exploring the trade-off between fidelity- and time-optimal control of quantum unitary transformations. Phys. Rev. A **86,** 062309 (2012)

[15] Gottesman, D.: Fault-tolerant quantum computation with higher-dimensional systems. Lect. Notes. Comput. Sci. **1509,** 302 (1999)

[16] Muthukrishnan, A., Stroud, C.R.: Multivalued logic gates for quantum computation. J. Mod. Optics **49,** 2115 (2002)

[17] Daboul, J., Wang, X., Sanders, B.C.: Quantum gates on hybrid qudits. J. Phys. A **36,** 2525 (2003)

[18] Greentree, A.D., Schirmer, S.G., Green, F., Hollenberg, L.C.L., Hamilton, A.R., Clark, R.G.: Maximizing the Hilbert space for a finite number of distinguishable quantum states. Phys. Rev. Lett. **92,** 097901 (2004)

[19] Bechmann-Pasquinucci, H., Pers, A.: Quantum cryptography with 3-state systems. Phys. Rev. Lett. **85,** 3313 (2002)

[20] Zobov, V.E., Pekhterev, D.I.: Adder on ternary base elements for a quantum computer. Pis'ma Zh. Eksp. Teor. Fiz. **89,** 303 (2009) [JETP Lett. **89,** 260 (2009)]

[21] Di, Y.-M., Wei, H.–R.: Elementary gates of ternary quantum logic circuit. ArXiv 1105.5485 (2013)

[22] Klimov, A.B., Guzman, R., Retamal, J.C., Saavedra, C.: Qutrit quantum computer with trapped ions. Phys.Rev. A **67,** 062313 (2003)

[23] Das, R., Mitra, A., Kumar, V., Kumar, A.: Quantum Information processing by NMR: Preparation of pseudopure states and implementation of unitary operations in a single-qutrit system. Int. J. Quantum Inf. **1,** 387 (2003)

[24] Shauro, V. P., Pekhterev, D.I., Zobov, V.E.: A comparative analysis of two methods of realizing elementary logic operators for a quantum computer on qutrits. Izv. Vyssh. Uchebn. Zaved. Fiz. **6,** 41 (2007 ) [Russ. Phys. J. **50,** 566(2007)].

[25] Vitanov, N.V.: Synthesis of arbitrary SU(3) transformations of atomic qutrits. Phys. Rev. A **85,** 032331 (2012)

[26] Dogra, S., Arvind, Dorai, K.: Determining the parity of a permutation using an experimental NMR qutrit. ArXiv 1402.5026 (2014)

[27] Ernst, R.R., Bodenhausen, D., Wokaun, A.: Principles of Nuclear Magnetic Resonance in One and Two Dimensions. Oxford, Clarendon (1987)

[28] Schirmer, S. G., Fu, H., Solomon, A.I.: Complete controllability of quantum systems. Phys. Rev. A **63,** 063410 (2001)

[29] Li, B., Yu, Z.-H., Fei, S.-M.: Geometry of quantum computation with qutrits. Sci. Rep. **3,** 2594 (2013)

[30] Zobov, V.E., Shauro, V.P.: Selective control of the states of a three-level quadrupole nucleus by means of nonselective rf pulses. Pis'ma Zh. Eksp. Teor. Fiz. **86,** 260 (2007) [JETP Lett. **86,** 230 (2007)]

[31] Zobov, V.E., Shauro, V.P.: Selective control of the states of multilevel quantum systems using nonselective rotation operators. Zh. Eksp. Teor. Fiz. **135,** 10 (2009) [JETP **108,** 5 (2009)]

[32] Shauro, V. P., Zobov, V.E.: Global phase and minimum time of quantum Fourier transform for qudits represented by quadrupole nuclei. Phys. Rev. A **88,** 042320 (2013)

[33] Zobov, V.E., Shauro, V.P.: Effect of a phase factor on the minimum time of a quantum gate. Zh. Eksp. Teor. Fiz. **145,** 25 (2014 ) [JETP **118,** 18 (2014)]

[34] Zobov, V.E., Shauro, V.P.: On time-optimal NMR control of states of qutrits represented by quadrupole nuclei with the spin I=1. Zh. Eksp. Teor. Fiz. **140,** 211 (2011) [JETP **113,** 181(2011)]